\documentclass[10pt]{article}
\usepackage{graphicx,epsfig,pstricks,rotating}
\usepackage{bm}

\newcommand{\eqnb}{\begin{equation}}
\newcommand{\eqne}{\end{equation}}

\newcommand{\question}[1]{}
\newcommand{\lsim}[1]{
\setlength{\unitlength}{12pt}
\begin{picture}(1.4,1.)
\put(.7,-0.3){\makebox(0.0,1.)[t]{$<$}}
\put(.7,-0.3){\makebox(0.0,1.)[b]{$\sim$}}
\end{picture}#1}
\newcommand{\NSt}{{\mbox{\scriptsize\it NS}}}
\newcommand{\St}{{\mbox{\scriptsize\it S}}}

\def\slashchar#1{\setbox0=\hbox{$#1$}  
   \dimen0=\wd0     
   \setbox1=\hbox{/} \dimen1=\wd1  
   \ifdim\dimen0>\dimen1   
      \rlap{\hbox to \dimen0{\hfil/\hfil}} 
      #1     
   \else     
      \rlap{\hbox to \dimen1{\hfil$#1$\hfil}} 
      /      
   \fi}      %

\begin{document}
\textwidth=135mm
\textheight=200mm

\begin{center}
{ \bfseries Pseudoscalar Meson Nonet at Zero and Finite 
Temperature\footnote{{\small Talk 
presented by D. Klabu\v{c}ar at the ``Dense Matter In Heavy Ion Collisions and 
Astrophysics'', JINR, Dubna,  August 21 -- September 1, 2006.}}}

\vskip 5mm

D. Horvati\' c$^{*,}\footnote{{\ttfamily davorh@phy.hr}}$,
D. Blaschke$^{\dag,\ddag,}\footnote{{\ttfamily david@theor.jinr.ru}}$,
D. Klabu\v{c}ar$^{*,}\footnote{{\ttfamily klabucar@phy.hr}, 
senior associate of Abdus Salam ICTP}$,
A. E. Radzhabov$^{\ddag,}\footnote{{\ttfamily aradzh@theor.jinr.ru}}$,

\vskip 5mm

{\small {\it $^*$ Physics Department, Faculty of Science,
University of Zagreb,\\
Bijeni\v{c}ka c. 32, Zagreb 10000, Croatia}}
\\
{\small {\it 
$^\dag$ 
Institut f\"ur Physik, Universit\"at Rostock
D-18051 Rostock, Germany
}}
\\
{\small {\it $^\ddag$ 
Bogoliubov Laboratory of Theoretical Physics,\\ 
Joint Institute for Nuclear Research,
141980 Dubna, Russia}}
\end{center}

\begin{abstract}
\noindent
Theoretical understanding of experimental results 
from relativistic heavy-ion collisions requires 
a microscopic approach to the behavior of QCD n-point functions
at finite temperatures, as given by the hierarchy of
Dyson-Schwinger equations, properly generalized within the
Matsubara formalism. The convergence of sums over Matsubara modes
is studied. The technical complexity of finite-temperature 
calculations mandates modeling. We present a model where the 
QCD interaction in the infrared, nonperturbative 
domain, is modeled by a separable form. Results 
for the mass spectrum of light quark flavors ($u$, $d$, $s$) 
and for the pseudoscalar bound-state amplitudes at 
finite temperature are presented.
\end{abstract}

\vskip 5mm

\section{Introduction}

While the experiments at RHIC \cite{Muller:2006ee,Adams:2005dq}
advanced the empirical knowledge of the hot QCD matter dramatically, 
the understanding of the state of matter that has been formed is 
still lacking.
For example, the STAR collaboration's assessment \cite{Adams:2005dq} 
of the evidence from RHIC experiments depicts a very intricate, 
difficult-to-understand picture of the hot QCD matter. 
Among the issues pointed out as important was the need
to clarify the role of quark-antiquark ($q\bar q$) bound states
continuing existence above the critical temperature $T_c$, as well
as the role of the chiral phase transition.

\begin{figure}
\centerline{\includegraphics[width=100mm,angle=0]{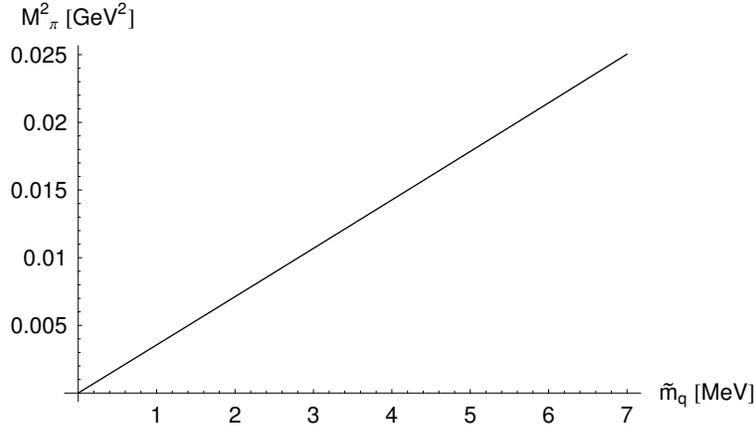}}
\caption{Correct chiral behavior of the pion mass close to chiral limit: 
$M_\pi^2 \propto {\widetilde m}_q $, where ${\widetilde m}_q $ is the 
bare quark mass parameter. It is akin to the notion of the current quark mass
in QCD and gives the extent of the explicit chiral symmetry breaking
as opposed to DChSB -- see Eq. (\ref{DSE}) below.
 }
\label{Mpi2}
\end{figure}

Both of these issues are consistently treated within
the Dyson-Schwinger (DS) approach to quark-hadron physics.
Dynamical chiral symmetry breaking (DChSB) as the crucial low-energy 
QCD phenomenon is well-understood in the rainbow-ladder approximation (RLA),
a symmetry preserving truncation of the hierarchy of DS equations.
Thanks to this, the behavior of the pion mass is in accord with the 
Goldstone theorem: the pion mass shows correct behavior while approaching 
the chiral limit, as seen on Fig.~\ref{Mpi2}. This correct chiral 
behavior is a general feature of the DS approach in RLA, and 
not a consequence of our specific model choice.
For recent reviews of the DS approach, see, e.g., Refs. 
\cite{Roberts:2000aa,Alkofer:2000wg}, of which the first 
\cite{Roberts:2000aa} also reviews the studies of
QCD  DS equations at finite temperature, started in \cite{Bender:1996bm}.
Unfortunately, the extension of DS calculations to non-vanishing 
temperatures is technically quite difficult. This usage of separable 
model interactions greatly simplifies DS calculations at finite temperatures, 
while yielding equivalent results on a given level of truncation  
\cite{Burden:1996nh,Blaschke:2000gd}. 
A recent update of this covariant separable approach with application to
the scalar $\sigma$ meson at finite temperature can be found in 
\cite{Kalinovsky:2005kx}.
Here, we present results for the quark mass spectrum
at zero and finite temperature, extending previous work by
including the strange flavor.


\section{The separable model at zero temperature}
\label{Model}

The dressed quark propagator $S_q(p)$ is the solution of its 
DS equation \cite{Roberts:2000aa,Alkofer:2000wg}, 
\begin{eqnarray}\label{sde}
S_q(p)^{-1} = i \gamma\cdot{p} + \widetilde{m}_q +
\frac{4}{3} \int \frac{d^4\ell}{(2\pi)^4} \, 
g^2 D_{\mu\nu}^{\mathrm{eff}} (p-\ell)
  \gamma_\mu S_q(\ell) \gamma_\nu \, , 
\label{DSE} 
\end{eqnarray}
while the $q\bar q'$ meson Bethe-Salpeter (BS) bound-state vertex 
$\Gamma_{q\bar q'}(p,P)$ is the solution of the BS equation (BSE)
\begin{eqnarray}\label{bse}
-\lambda(P^2)\Gamma_{q\bar q'}(p,P) = \frac{4}{3} \int \frac{d^4\ell}{(2\pi)^4}  
 g^2  D_{\mu\nu}^{\mathrm{eff}} (p-\ell)
\gamma_\mu S_q(\ell_+) \Gamma_{q\bar q'}(\ell,P) S_q(\ell_-) \gamma_\nu, \, 
\label{BSE}
\end{eqnarray}
where $D_{\mu\nu}^{\mathrm{eff}}(p-\ell)$ is an effective gluon propagator 
modeling the nonperturbative QCD effects, 
$\widetilde{m}_q$ is the current quark mass, the index $q$ (or $q'$) stands for
the quark flavor ($u, d$ or $s$),
$P$ is the total momentum, and $\ell_{\pm}=\ell\pm P/2$. 
The chiral limit is obtained by setting $\widetilde{m}_q=0$.
The meson mass is identified from $\lambda(P^2=-M^2)=1$.
Equations (\ref{DSE}) and (\ref{BSE}) are written in the Euclidean space,
and in the consistent rainbow-ladder truncation.

\begin{figure}[!hbt]
\centerline{\includegraphics[width=120mm,angle=0]{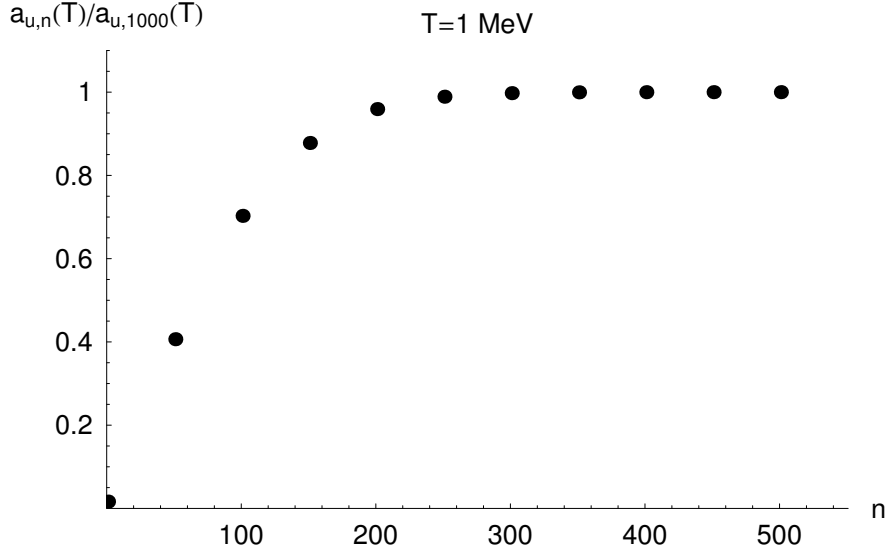}}
\caption{The sum in Eq.~(\ref{gapat}) as the function of number of the
Matsubara modes included in summation at temperature $T=1$ MeV.
Eq.~(\ref{gapat}) is normalized to value calculated with enough
Matsubara modes ($n=1000$) to achieve prescribed numerical precision.}
\label{mats1}
\end{figure}

The simplest separable Ansatz 
which reproduces in RLA a nonperturbative solution of (\ref{DSE}) for any 
effective gluon propagator in a Feynman-like gauge
$g^2 D_{\mu\nu}^{\mathrm{eff}} (p-\ell) \rightarrow
\delta_{\mu\nu} D(p^2,\ell^2,p\cdot \ell)$ is 
\cite{Burden:1996nh,Blaschke:2000gd}
\begin{eqnarray}
D(p^2,\ell^2,p\cdot \ell)=D_0 {\cal F}_0(p^2) {\cal F}_0(\ell^2) 
+ D_1 {\cal F}_1(p^2) (p\cdot \ell ) {\cal F}_1(\ell^2)~.
\label{sepAnsatz}
\end{eqnarray}
This is a rank-2 separable interaction with two strength parameters $D_i$ 
and corresponding form factors ${\cal F}_i(p^2)$, $i=1,2$. 
The choice for these quantities is constrained to the solution of
the DSE for the quark propagator (\ref{DSE})
\begin{eqnarray}
  S_q(p)^{-1} = i \gamma\cdot{p} A_q(p^2) + B_q(p^2) 
\equiv Z^{-1}_q(p^2) [ i \gamma\cdot{p} + m_q(p^2) ]~,
\end{eqnarray}
where $m_q(p^2)=B_q(p^2)/A_q(p^2)$ is the dynamical mass function and
$Z_q(p^2)=A^{-1}_q(p^2)$ the wave function renormalization.
Using the separable Ansatz (\ref{sepAnsatz}) in (\ref{sde}), 
the gap equations for the quark amplitudes $A_q(p^2)$ and $B_q(p^2)$ read
\begin{eqnarray}
B_q(p^2) - \widetilde{m}_q =  
\frac{16}{3} \int \frac{d^4\ell}{(2\pi)^4} D(p^2,\ell^2,p \cdot \ell)
\frac{B_q(\ell^2)}{\ell^2 A_q^2(\ell^2)+ B_q^2(\ell^2)} 
=  b_q {\cal F}_0(p^2)\, ,
\label{gap1}\\
\left[A_q(p^2)-1 \right]  
=\frac{8}{3p^2} \int \frac{d^4\ell}{(2\pi)^4} D(p^2,\ell^2,p \cdot \ell)
\frac{(p\cdot \ell) A_q(\ell^2)}{\ell^2A_q^2(\ell^2)+B_q^2(\ell^2)} 
= a_q {\cal F}_1(p^2)\, .
\label{gap2}
\end{eqnarray}
Once the coefficients $a_q$ and $b_q$ are
obtained by solving the gap equations (\ref{gap1}) and (\ref{gap2}), 
the only model parameters remaining are $\widetilde{m}_q$ and the parameters 
of the gluon propagator, to be fixed by meson phenomenology.

\begin{figure}[!hbt]
\centerline{\includegraphics[width=120mm,angle=0]{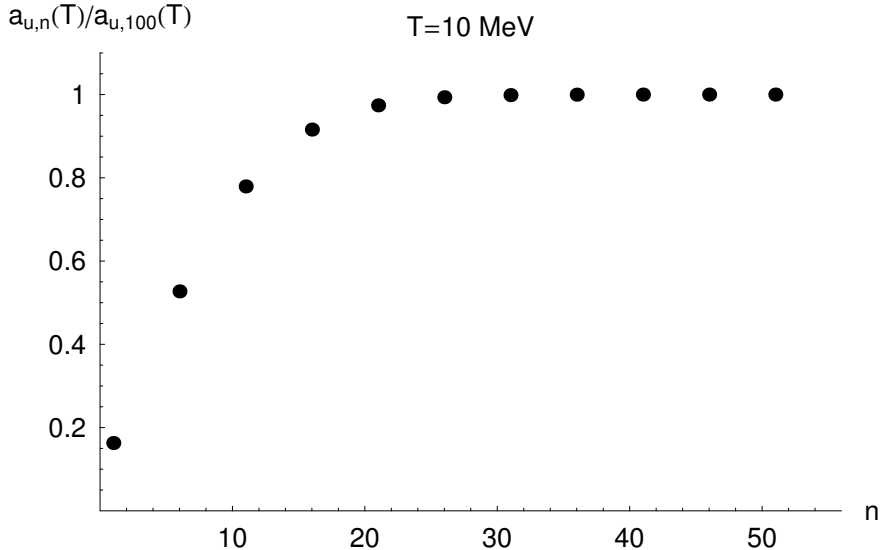}}
\caption{The sum in Eq.~(\ref{gapat}) as the function of number of
the Matsubara modes included in summation at temperature $T=10$ MeV.
Eq.~(\ref{gapat}) is normalized to value calculated with enough
Matsubara modes ($n=100$) to achieve prescribed numerical precision.}
\label{mats10}
\end{figure}

\begin{figure}[!hbt]
\centerline{\includegraphics[width=120mm,angle=0]{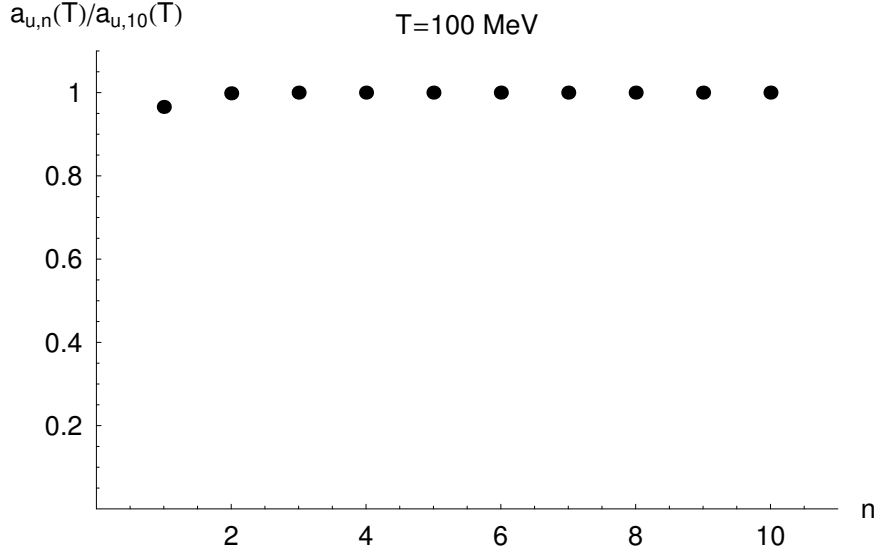}}
\caption{The sum in the Eq.~(\ref{gapat}) as function of number of
the Matsubara modes included in summation at
temperature $T=100$ MeV. Eq.~(\ref{gapat}) is normalized to value
calculated with enough Matsubara modes ($n=10$) to
achieve prescribed numerical precision.}
\label{mats100}
\end{figure}

\section{Extension to finite temperature}

The extension of the separable model studies to the finite temperature case, 
$T\neq 0$, is systematically accomplished by a transcription of the Euclidean 
quark 4-momentum via {$p \rightarrow$} {$ p_n =$} {$(\omega_n, \vec{p})$}, 
where {$\omega_n=(2n+1)\pi T$} are the
discrete Matsubara frequencies. In the Matsubara formalism, the number of
coupled equations represented by (\ref{sde}) and (\ref{bse}) scales up
with the number of fermion Matsubara modes included.  For studies near
and above the transition, \mbox{$T \ge 100 $}~MeV, using only 10 such modes
appears adequate. Nevertheless, the appropriate number can be more than 
$10^3$ if the continuity with \mbox{$T=0$} results is to be verified. 
Convergence of the sum in the Eq.~(\ref{gapat}) is shown on 
Figs.~\ref{mats1}--\ref{mats100}. The effective $\bar q q$ interaction,  
defined in the present paper by the Ansatz (\ref{sepAnsatz}) and the 
form factors (\ref{model_interaction}) below, will automatically 
decrease with increasing $T$ without the introduction of an explicit 
$T$-dependence which would require new parameters.

The solution of the DS equation for the dressed quark propagator 
now takes the form
\begin{eqnarray}
S_q^{-1}(p_n, T) = i\vec{\gamma} \cdot \vec{p}\; A_q(p_n^2,T)
                   + i \gamma_4 \omega_n\; C_q(p_n^2,T)+ B_q(p_n^2,T),\;
\label{invprop}
\end{eqnarray}
where {$p_n^2=\omega_n^2 + \vec{p}^{\,2}$}
and the quark amplitudes 
{$B_q(p_n^2,T) = \widetilde{m}_q + b_q(T) {\cal F}_0(p_n^2)$}, 
\mbox{$A_q(p_n^2,T) = 1+ a_q(T) {\cal F}_1(p_n^2)$}, and 
{$C_q(p_n^2,T) = 1+ c_q(T) {\cal F}_1(p_n^2)$}
are defined by the temperature-dependent coefficients 
$a_q(T)$,$ b_q(T)$, and $c_q(T)$ to be determined from the set of 
three coupled non-linear equations
\begin{eqnarray}
a_q(T) = \frac{8 D_1}{9}\,  T \sum_n \int \frac{d^3p}{(2\pi)^3}\,{\cal F}_1(p_n^2)\, \vec{p}^{\,2}\, [1 + a_q(T) {\cal F}_1(p_n^2)]\; d_q^{-1}(p_n^2,T) \; , \label{gapat}
\\
 c_q(T) = \frac{8 D_1}{3}\,  T \sum_n \int \frac{d^3p}{(2\pi)^3}\,{\cal F}_1(p_n^2)\, \omega_n^2\, [1 +  c_q(T) {\cal F}_1(p_n^2)]\;
                                                d_q^{-1}(p_n^2,T) \; ,
\\
 b_q(T) = \frac{16 D_0}{3}\,  T \sum_n \int \frac{d^3p}{(2\pi)^3}\,{\cal F}_0(p_n^2)\, [\widetilde{m}_q +  b_q(T) {\cal F}_0(p_n^2)]\; d_q^{-1}(p_n^2,T) \; .
\end{eqnarray}
The function $d_q(p_n^2,T)$ is the denominator of the quark 
propagator $S_q(p_n, T)$, and is given by
\begin{eqnarray}
d_q(p_n^2,T) = \vec{p}^{\,2}A_q^2(p_n^2,T) +\omega_n^2C_q^2(p_n^2,T)
                  + B_q^2(p_n^2,T).
\label{Sdenominator}
\end{eqnarray}

The procedure for solving gap equations for a given temperature $T$ is the
same as in $T=0$ case, but one has to control the appropriate
number of Matsubara modes as mentioned above.

\begin{figure}[!hbt]
\centerline{\includegraphics[width=120mm,angle=0]{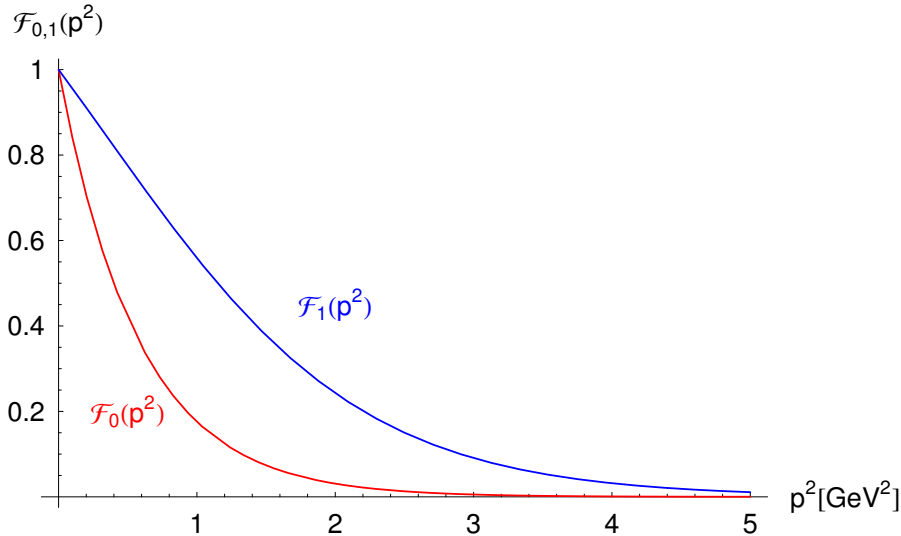}}
\caption{The $p^2$ dependence of the form factors ${\cal F}_0(p^2)$ and ${\cal F}_1(p^2)$.}
\label{figformf}
\end{figure}

\section{Confinement and Dynamical Chiral Symmetry Breaking}

If there are no poles in the quark propagator $S_q(p)$ for real timelike
$p^2$ then there is no physical quark mass shell. This entails 
that quarks cannot propagate freely, and the description of hadronic
processes will not be hindered by unphysical quark production thresholds.
This sufficient condition is a viable possibility for realizing quark 
confinement~\cite{Blaschke:2000gd}. A nontrivial solution for $B_q(p^2)$ 
in the chiral limit (${\widetilde m}_0=0$) signals DChSB.
There is a connection between quark confinement realized as the lack 
of a quark mass shell and the existence of a strong quark mass function 
in the infrared through DChSB.
The propagator is confining if \mbox{$m_q^2(p^2) \neq -p^2$} for real $p^2$,
where the quark mass function is \mbox{$m_q(p^2)=B_q(p^2)/A_q(p^2)$}.
In the present separable model, the strength
\mbox{$b_q=B_q(0)$}, which is generated by solving Eqs. (\ref{gap1})
and (\ref{gap2}), controls both confinement and DChSB.
At finite temperature, the strength $b_q(T)$ for the quark mass function 
will decrease with $T$, until the denominator (\ref{Sdenominator}) of the
quark propagator can vanish for some timelike momentum, and the quark can 
come on the free mass shell. The connection between deconfinement and
disappearance of DChSB is thus clear in the DS approach. Also the
present model is therefore expected to have a deconfinement transition 
at or a little before the chiral restoration transition associated with 
\mbox{$b_0(T) \to 0$}.

The following simple choice of the separable interaction form factors (graphically 
represented on Fig.~\ref{figformf}),
\begin{equation}
{\cal F}_0(p^2)=\exp(-p^2/\Lambda_0^2)~,~~
{\cal F}_1(p^2)=\frac{1+\exp(-p_0^2/\Lambda_1^2)}
{1+\exp((p^2-p_0^2)/\Lambda_1^2)}~,
\label{model_interaction}
\end{equation}
is used to obtain numerical solutions which reproduce
very well the phenomenology of the light pseudoscalar mesons and 
generate an acceptable value for the chiral condensate.

The resulting quark propagator is found to be confining and the
infrared strength and shape of quark amplitudes
$A_q(p^2)$ and $B_q(p^2)$ are in quantitative agreement with
the typical DS studies. Gaussian-type form factors are used
as a minimal way to preserve these properties while realizing
that the ultraviolet suppression is much stronger than the asymptotic
fall off (with logarithmic corrections) known from perturbative QCD
and numerical studies on the lattice \cite{Parappilly:2005ei}.

\begin{figure}[!hbt]
\centerline{\includegraphics[width=120mm,angle=0]{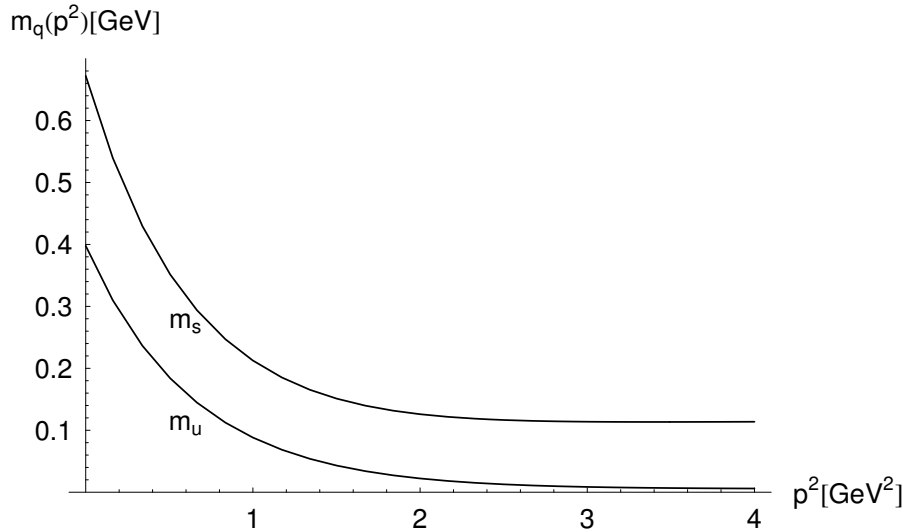}}
\caption{The $p^2$ dependence (at $T=0$) of the dynamically
generated quark masses $m_s(p^2), m_u(p^2)$ for, respectively,
the strange and the (isosymmetric) nonstrange case.}
\label{figMp2}
\end{figure}

\section{Bound-state amplitudes}

With the separable interaction, the allowed form of the solution 
of Eq. (\ref{bse}) for the pseudoscalar BS amplitude
is \cite{Burden:1996nh}
\begin{eqnarray}
\Gamma_{PS}(\ell;P) =\gamma_5 \left(i E_{PS} (P^2) +
                     \slashchar{P} F_{PS}(P^2)\right) \; {\cal F}_0(\ell^2).
                     \label{psbsa}
\end{eqnarray}
The dependence on the relative momentum $\ell$ is described only by the first 
form factor ${\cal F}_0(\ell^2)$. The second term ${\cal F}_1$ of the interaction 
can contribute to BS amplitude only indirectly via the quark propagators. 
The pseudoscalar BSE (\ref{bse}) becomes a $2\times 2$ matrix 
eigenvalue problem \mbox{${\cal K}(P^2) {\cal V} = \lambda(P^2) {\cal V}$} 
where the eigenvector is \mbox{${\cal V} = (E_{PS}, F_{PS})$}. The kernel is
\begin{eqnarray}
{\cal K}_{ij}(P^2) = - \frac{4 D_0}{3}\, {\rm tr_s}\, \int\,
       \frac{d^4\ell}{(2\pi)^4}{\cal F}_0^2(\ell^2)
 \left[ \hat{t}_i\, S_q(\ell_+)\, t_j\, S_{\bar{q}}(\ell_-)\,  \right]~,
\label{pskernel}
\end{eqnarray}
where $i,j=1,2$ denote the components of 
\mbox{$t=(i\gamma_5, \gamma_5\, \slashchar{P})$} and
\mbox{$\hat{t}=(i\gamma_5,$} \mbox{$-\gamma_5\, \slashchar{P}/2P^2)$}. 
The separable model produces the same momentum dependence
for both amplitudes (containing $F_{PS}$ and $E_{PS}$) in the 
BS amplitude (\ref{psbsa}): the dependence of the quark amplitude 
$B_q(\ell^2)$. Goldstone's theorem is preserved by the present
separable model; in the chiral limit, whenever a nontrivial 
gap-equation solution for $B_q(p^2)$ exists, there will be 
a massless pion solution to (\ref{pskernel}).

The normalization condition for the pseudoscalar BS amplitude can be expressed as
\begin{eqnarray}
\label{pinorm}
 2 P_\mu &=& N_f N_c \,  \frac{\partial}{\partial P_\mu} \,
        \,  \int \frac{d^4\ell}{(2\pi)^4}\,
        {\rm tr}_s \left[
        \bar\Gamma_{PS}(\ell;-K)\, \right.\times \nonumber \\
         &\times&\left.\left.S_q(\ell_+)\,
        \Gamma_{PS}(\ell;K)\,S_{\bar{q}}(\ell_-) \right]
        \right|_{P^2=K^2=-M_{PS}^2}.
\end{eqnarray}

Here $\bar\Gamma(q;K)$ is the charge conjugate amplitude $[{\cal C}^{-1} \Gamma(-q,K) {\cal
C}]^{\rm t}$,  where ${\cal C}=\gamma_2 \gamma_4$ and the index t denotes a matrix transpose.
Both the number of colors $N_c$ and light flavors $N_f$ are 3.

At \mbox{$T=0$} the mass-shell condition for a meson as a $\bar q q$ bound state of the BSE
is equivalent to the appearance of a pole in the $\bar q q$ scattering amplitude as a
function of $P^2$.  At $T\neq0$ in the Matsubara formalism, the $O(4)$ symmetry is broken
by the heat bath and we have \mbox{$P \to (\Omega_m,\vec{P})$} where \mbox{$\Omega_m = 2m
\pi T$}.  Bound states and the poles they generate in propagators may be investigated
through polarization tensors, correlators or Bethe-Salpeter eigenvalues.  This pole
structure is characterized by information at discrete points $\Omega_m$ on the imaginary
energy axis and at a continuum of 3-momenta. One may search for poles as a function of
$\vec{P}^2$ thus identifying the so - called spatial or screening masses for each Matsubara
mode.  These serve as one particular characterization of the propagator and the \mbox{$T >
0 $} bound states.

In the present context, the eigenvalues of the BSE become $\lambda(P^2) \to
\tilde{\lambda}(\Omega_m^2,\vec{P}^2;T)$. The temporal meson masses identified by zeros of
$1-\tilde{\lambda}(\Omega^2,0;T)$ will differ in general from the spatial masses identified
by zeros of $1-\tilde{\lambda}(0,\vec{P}^2;T)$.  They are however identical at \mbox{$T =0$}
and an approximate degeneracy can be expected to extend over the finite $T$ domain, where the
$O(4)$ symmetry is not strongly broken.

The general form of the finite-$T$ pseudoscalar BS amplitude allowed by the separable model is
\begin{eqnarray}
\Gamma_{PS}(q_n;P_m) &=&\gamma_5 \left(i E_{PS} (P_m^2)
  +  \gamma_4 \, \Omega_m \tilde{F}_{PS}(P_m^2)\right. \nonumber \\
  &+&  \left. \vec{\gamma} \cdot \vec{P} F_{PS}(P_m^2)\right) \; {\cal F}_0(q_n^2) \; .
\label{pibsaT}
\end{eqnarray}
The separable BSE becomes a $3\times 3$ matrix eigenvalue problem with a kernel that is a
generalization of Eq.~(\ref{pskernel}).  In the limit \mbox{$\Omega_m \to 0$}, as is
required for the spatial mode of interest here, the amplitude \mbox{$\hat{F}_{PS} = \Omega_m
\tilde{F}_{PS}$} is trivially zero.

\section{Results}

Parameters of the model are completely fixed by meson phenomenology calculated
from the model as discussed in \cite{Blaschke:2000gd,Kalinovsky:2005kx}. 
In the nonstrange sector, we work in the isosymmetric limit and 
adopt bare quark masses ${\widetilde m}_u = {\widetilde m}_d = 5.5$ MeV and
in strange sector we adopt ${\widetilde m}_s = 115$ MeV.
Then the parameter values 
\begin{equation}
\Lambda_0=758 \, {\rm MeV}, \quad 
\Lambda_1=961 \, {\rm MeV}, \quad 
p_0=600 \, {\rm MeV},
\label{Lambda12p0}
\end{equation}
\begin{equation}
D_0\Lambda_0^2=219 \, , \qquad D_1\Lambda_1^4=40 \, ,
\label{D0D1}
\end{equation} 
lead, through the gap equation, to 
$a_{u,d}=0.672$, $b_{u,d}=660$ MeV, $a_{s}=0.657$ and $b_{s}=998$ MeV 
i.e., to the dynamically generated momentum-dependent mass functions
$m_q(p^2)$ shown in Fig. \ref{figMp2}. 
In the limit of high $p^2$, they converge to ${\widetilde m}_u$ and 
${\widetilde m}_s$.  However, at low $p^2$, the values of 
$m_u(p^2)$ are close to the typical constituent quark 
mass scale $\sim m_\rho/2 \sim m_N/3$ with the maximum value at $p^2=0$, 
$m_u(0)=398$ MeV. 
The corresponding value for the strange quark is $m_s(0)=672$ MeV.
Fig. \ref{figMp2} hence shows that in the domain of low
and intermediate $p^2 \lsim 1$ GeV$^2$, the dynamically
generated quark masses are of the order of typical 
constituent quark mass values. 

Thus, the DS approach provides a derivation of the 
constituent quark model \cite{Kekez:1998xr} from a more fundamental level, 
with the correct chiral behavior of QCD.

\begin{figure}[!hbt]
\centerline{\includegraphics[width=120mm,angle=0]{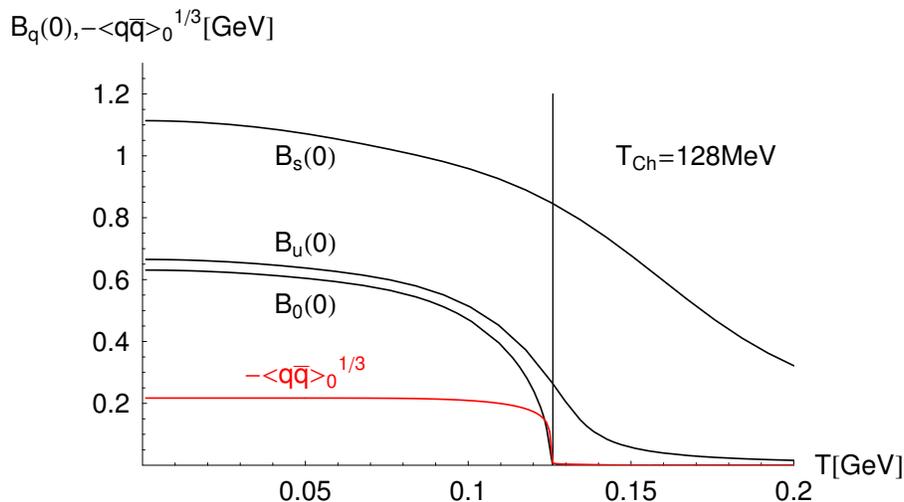}}
\caption{The temperature dependence of $B_s(0), B_u(0)$ and $B_0(0)$,
the scalar propagator functions at $p^2=0$, for
the strange, the nonstrange and the chiral-limit cases, respectively.
The temperature dependence of the chiral quark-antiquark condensate,
$-\langle q\bar{q}\rangle^{1/3}_0$, is also shown (by the
lowest curve). Both chiral-limit quantities,
$B_0(0)$ and $-\langle q\bar{q}\rangle^{1/3}_0$, vanish at the
chiral-symmetry restoration temperature $T_\mathrm{Ch}=128$ MeV.
}
\label{figBT}
\end{figure}

Obtaining such dynamically generated constituent quark masses, 
as previous experience with the DS approach shows (see, e.g.,
Refs. such as
\cite{Roberts:2000aa,Alkofer:2000wg,Kekez:1998xr
}),
is essential for reproducing the static
and other low-energy properties of hadrons, including decays.
(We would have to turn to less simplified DS models for incorporating
the correct perturbative behaviors, including that of the quark masses.
Such models are amply reviewed or used in, e.g., Refs.
\cite{Roberts:2000aa,Alkofer:2000wg,Kekez:1998xr,
Klabucar:1997zi},
but addressing them is beyond the present scope, where the
perturbative regime is not important.)

Another important result related to the dynamically generated,
dressed quark propagator, is the chiral quark-antiquark condensate
$\langle q\bar{q}\rangle_0$. For the parameter values quoted above,
we obtain the zero-temperature value 
$\langle q\bar{q}\rangle_0 = (-217 \, {\rm MeV})^3$, which
practically coincides with the standard QCD value.

\begin{figure}[!hbt]
\centerline{\includegraphics[width=120mm,angle=0]{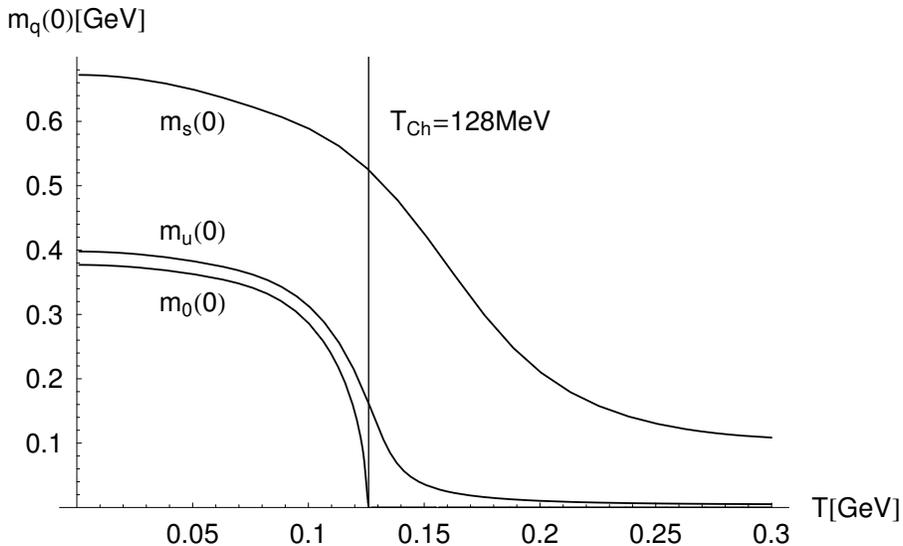}}
\caption{The temperature dependence of $m_s(0), m_u(0)$ and $m_0(0)$, 
the dynamically generated quark masses at $p^2=0$ for 
the strange, the nonstrange and the chiral-limit cases, respectively.}
\label{figMT}
\end{figure}

The extension of these results to finite temperatures is given in 
Figs. \ref{figBT}, \ref{figMT}. Very important is the 
temperature dependence of the chiral-limit quantities $B_0(0)_T$ 
and $\langle q \bar q \rangle_{0}(T)$, whose vanishing with $T$ 
determines the chiral restoration temperature $T_\mathrm{Ch}$.
We find $T_\mathrm{Ch} = 128$ MeV in the present model.


\begin{figure}[!hbt]
\centerline{\includegraphics[width=120mm,angle=0]{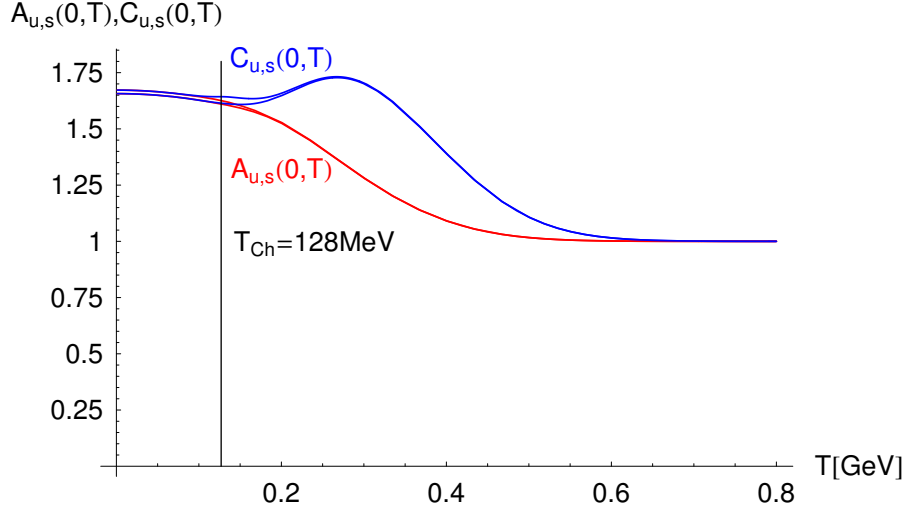}}
\caption{The violation of $O(4)$ symmetry with $T$ is exhibited 
on the example of $A_{u,s}(0,T)$ and $C_{u,s}(0,T)$.}
\label{figACT}
\end{figure}

The temperature dependences of the functions giving the vector part 
of the quark propagator, $A_{u,s}(0)_T$ and $C_{u,s}(0)_T$,
are depicted in Fig. \ref{figACT}. Their difference is a measure
of the O(4) symmetry breaking with the temperature $T$.

\begin{figure}[!hbt]
\centerline{\includegraphics[width=120mm,angle=0]{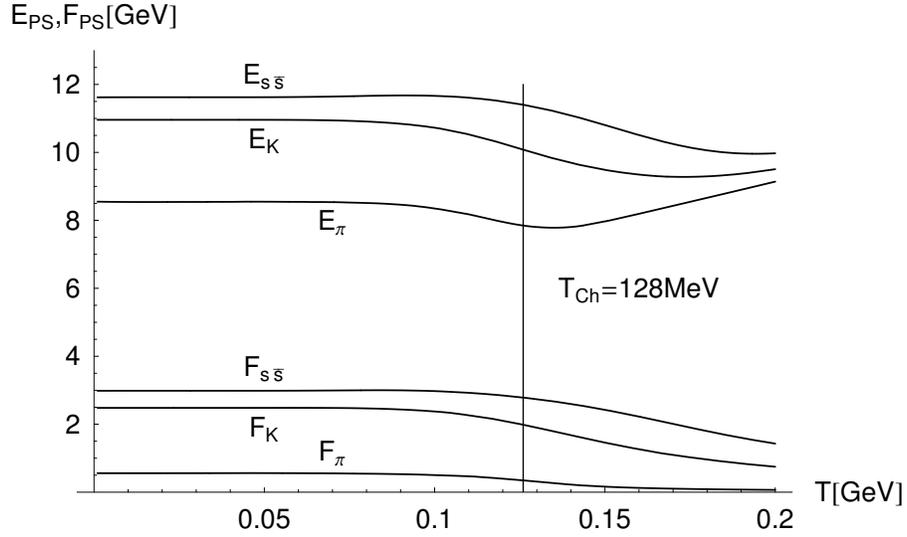}}
\caption{The temperature dependence of pseudoscalar covariants in BS amplitude}
\label{figvert}
\end{figure}

The results for pseudoscalar $E_{PS}$ and pseudovector $F_{PS}$ amplitudes can be
seen on Fig.~\ref{figvert}. The pseudovector amplitude for pion $F_{\pi}$ is significantly different
from zero but decreases rapidly above the transition. 

The presented model, when applied in the framework of the
Bethe-Salpeter approach to mesons as 
quark-antiquark bound states, produces a very satisfactory 
description of the whole light pseudoscalar nonet, both 
at zero and finite temperatures \cite{Blaschke+alZTF}. 
The masses and decay constants of the pseudoscalar nonet
mesons at $T=0$ are summarized and compared with experiment
in Table \ref{piKssbarTable}.

The first three rows in Table \ref{piKssbarTable} give the 
masses $M_{PS}$ and decay constants $f_{PS}$ of the pseudoscalar
$q{\bar q}'$ bound states ${PS} = \pi^+, K^+$, and $s\bar s$
resulting, through the BS equation (\ref{BSE}),
from the separable interaction (\ref{sepAnsatz}).
The $s\bar s$ pseudoscalar meson is a useful
theoretical construct, but is not realized
physically, at least not at $T=0$. (It is therefore not
associated with any experimental value in this table.
Also note that the unphysical mass $M_{s\bar s}$ given
in this table does not include the contribution from the
gluon anomaly.)
The parameter values [(\ref{Lambda12p0}) and (\ref{D0D1}) in
the effective interaction $D_{\mu\nu}^{\rm eff}$,
and the bare quark masses ${\widetilde m}_{u,d} = 5.5$ MeV
and ${\widetilde m}_s = 115$ MeV] are fixed by fitting
the pion and kaon masses and decay constants.
These masses and decay constants are the input for
the description of the $\eta$--$\eta^\prime$ complex
\cite{Klabucar:1997zi}.
More precisely, $\eta$ and $\eta'$ masses are obtained by 
combining the contributions from the non-Abelian (gluon)
axial anomaly with the non-anomalous contributions 
obtained from the results on the masses of 
$\pi, K$, and the unphysical $s\bar s$ pseudoscalar
\cite{Klabucar:1997zi}. For this procedure, it is essential
that we have the good chiral behavior of our $q{\bar q}'$
bound states, which are simultaneously also the 
(almost-)Goldstone bosons, so that 
\begin{equation}
M_{q{\bar q}'}^2 = {\rm Const} \, ({\widetilde m}_q + {\widetilde m}_{q'}) \, ,
\label{goodChiral}
\end{equation}
as seen in Fig. \ref{Mpi2}. For example, Eq. (\ref{goodChiral})
guarantees the relation $M_{\pi}^2 + M_{s\bar s}^2 = 2 M_K^2$
which is utilized in Refs. \cite{Klabucar:1997zi,Blaschke+alZTF}
in the treatment of the $\eta$-$\eta'$ complex. 
Indeed, the concrete model results for $M_{\pi}, M_K$ and
$M_{s\bar s}$ in Table \ref{piKssbarTable} 
obey this relation up to $\frac{1}{4}$\%.

\begin{figure}[!hbt]
\centerline{\includegraphics[width=120mm,angle=0]{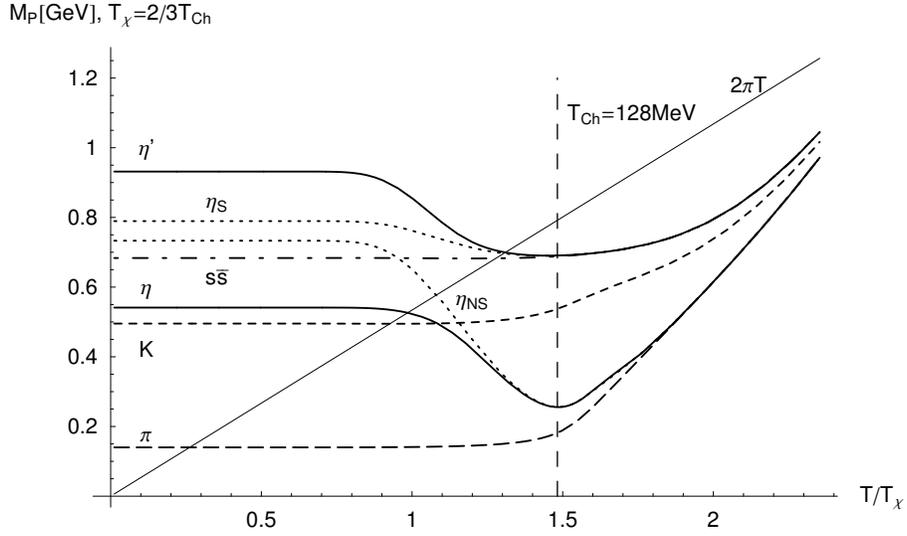}}
\caption{The relative temperature dependence, on $T/T_\chi$,
of the pseudoscalar meson masses for $T_\chi = 2/3 \, T_{\rm Ch}$.
The three variously dashed curves represent the pseudoscalar meson
masses which do not receive contributions from the gluon anomaly:
$M_{\pi}(T)$, $M_{K}(T)$ and $M_{s\bar s}(T)$,
depicted by the long-dashed, short dashed and dash-dotted
curves, respectively. The lower solid curve is $M_{\eta}(T)$,
and the upper solid curve is $M_{\eta'}(T)$.
The lower and  upper dotted curves are the masses
of $\eta_\NSt$ and $\eta_\St$. The thin diagonal line is
twice the zeroth Matsubara frequency, $2\pi T$. This is
the limit to which meson masses should ultimately approach
from below at still higher temperatures, where $q\bar q$ states
should totally dissolve into a gas of weakly interacting
quarks and antiquarks.
 }
\label{figPseudoTc85}
\end{figure}

\begin{table}[!ht]
\begin{center}
\begin{tabular}{|c|c|c|c|c|c|}
\hline
$ {PS} $   & $M_{PS}$ & $M_{PS}^{\rm exp}$ & $f_{PS}$ & $f_{PS}^{\rm exp}$
\\
\hline
  $\pi^+$     & {0.140} & {0.1396} & {0.092}  & ${0.0924\pm 0.0003}$ \\
\hline
  $K^+$       & {0.495} & {0.4937}  & {0.110} & ${0.1130\pm 0.0010}$ \\
\hline
  $s\bar s$ & {0.685} &          & {0.119} &    \\
\hline
  $\eta$    & {0.543} & 0.5473   &          &        \\
\hline
  $\eta'$   & {0.933} & 0.9578   &          &         \\
\hline
\end{tabular}
\end{center}
\caption{
Results on the pseudoscalar mesons at zero temperature, $T=0$, 
and comparison with experiment (where appropriate). 
All results are in GeV.
}
\label{piKssbarTable}
\end{table}
Especially interesting is the temperature behavior of the 
$\eta$-$\eta'$ complex, where the results for the meson 
masses differ very much for various possible relationships
between the chiral restoration temperature $T_\mathrm{Ch}$ 
and the temperature of melting of the topological susceptibility,
denoted by $T_\chi$. 
The once favored scenario of Pisarski and Wilczek \cite{Pisarski:ms},
where $\eta$ would smoothly evolve with $T$ to purely non-strange 
$\eta_\NSt$,
        \begin{equation}
 |\eta_\NSt\rangle =
        \frac{1}{\sqrt{2}} (|u\bar{u}\rangle + |d\bar{d}\rangle)
  = \frac{1}{\sqrt{3}} |\eta_8\rangle + \sqrt{\frac{2}{3}} |\eta_0\rangle~,
\label{etaNSdef}
\end{equation}
and $\eta'$ to purely strange $\eta_\St$, 
       \begin{equation}
 |\eta_\St\rangle =
            |s\bar{s}\rangle
  = - \sqrt{\frac{2}{3}} |\eta_8\rangle + \frac{1}{\sqrt{3}} |\eta_0\rangle~,
\label{etaSdef}
        \end{equation}
would occur only when  $T_\chi$ is significantly below $T_\mathrm{Ch}$.
Such a case is depicted in Fig. \ref{figPseudoTc85}, where 
$T_\chi = 2/3 \, T_\mathrm{Ch}$. In this case, around the 
chiral restoration temperature $\eta$ becomes quite light, 
and one would expect an increase of the relative multiplicity 
of $\eta$ mesons around $T_\mathrm{Ch}$. Nevertheless, 
the possibility that $T_\chi < T_\mathrm{Ch}$ is nowadays
disfavored by the lattice results on the temperature 
dependence of the topological susceptibility \cite{Alles:1996nm}.

On the other hand, for $T_\chi \sim T_\mathrm{Ch}$ 
or $T_\chi > T_\mathrm{Ch}$ we find that $\eta$ never
becomes light, while $\eta'$ even becomes very heavy
\cite{Blaschke+alZTF}.
Thus, for the relationships favored by the lattice, 
our results \cite{Blaschke+alZTF} indicate so strong
suppression of $\eta'$ around the chiral restoration 
temperature, that it may constitute a useful signal 
from the hot QCD matter.

\subsubsection*{Acknowledgments}

We thank M. Bhagwat, Yu.L. Kalinovsky and P.C. Tandy for discussions.
A.E.R.~acknowledges support by RFBR grant No. 05-02-16699, the
Heisenberg-Landau program and the HISS Dubna program of the Helmholtz 
Association. D.H.~and~D.K. were supported by MZT project No.~0119261.
D.B. is grateful for support by the Croatian Ministry of Science for a
series of guest lectures held in the Physics Department at University of
Zagreb, where the present work has been completed. D.K. acknowledges the
partial support of Abdus Salam ICTP at Trieste, where a part of this 
paper was written.

\end{document}